# Chemical Gradients in Galaxy Clusters and the Multiple Ways of Making a Cold Front

Renato A. Dupke, rdupkeumich.edu
Dept. Astronomy, University of Michigan

## 1 - Introduction

Cold fronts were originally interpreted as being the result of subsonic/transonic motions of head-on merging substructures with suppressed thermal conduction (Markevitch et al. 2000, 2001; Vikhlinin et al. 2001). This merger core remnant model is theoretically justified (e.g. Bialek & Evrard 2002; Nagai & Kravtsov 2003) and holds relatively well for clusters that have clear signs of merging, such as 1E0657-56 (Markevitch et al. 2002), but it does not work well for the increasing number of cold fronts found in well-behaved cold clusters such as, among others, A496 (Dupke & White 2003, hereafter DW03), A1795 (Markevitch et al. 2001), RXJ1720.1+2638 (Mazzotta et al. 2001). This prompted the re-evaluation of other models for cold front generation, such as oscillation of the cD+central gas (Fabian et al. 2001; DW03), hydrodynamic gas sloshing due to scattering of smaller halos (Tittley & Henriksen 2005; Ascasibar & Markevitch 2006, hereafter AM06).

As pointed out by DW03, different models for cold front formation can be discriminated through the analysis of SN Ia/II contamination across the front. If the cold front is generated by a merger core remnant we should expect the front to be accompanied by a characteristic discontinuity of metal enrichment histories, which can be determined through measurements of elemental abundance ratios. DW03 performed a chemical analysis of the cold front in A496. With an effective exposure of ∼9 ksec, they found no clear chemical discontinuities uniquely related to the cold front. Here we report the results of a deeper (effectively 55 ksec) observation of that cluster (Fig. 1a) that allowed us to produce high quality maps of the gas parameters and to compare more closely the observations with the predictions given by different models for cold front formation. We found for the first time a "cold arm"-cold front association characteristic of a flyby of a massive DM halo near the core of the cluster. The cold arm is accompanied by an enhanced SN II Fe mass fraction, inconsistent with the merger core remnant scenario. We assume $H_0 = 70$ km s$^{-1}$Mpc$^{-1}$, $\Omega_0 = 1$ and $1'' \approx 0.66$ kpc.

## 2 - Results

The temperature and abundance ratio maps (determined from a single absorbed VAPEC spectral model) are shown in Figs 1b,c, with X-ray surface brightness contours overlaid. This cluster shows multiple cold fronts (Dupke et al. 2006). One striking feature that can be seen in the temperature map is a "cold spiral arm" that departs from the core o the N-NW up to the cold front position and runs along the cold front to the E-NE, becoming more diffuse towards the S. The temperature of the arm is ≠3.0±0.2 keV. The temperatures on the surrounding regions of the cold arm are 3.8 keV and 4.3 keV towards the inner and outer cluster regions, respectively. This arm is definitely associated to the N (main) cold front and also to a lesser extent to the W cold front. The arm departs from a boxy low temperature region, the edges of which appear to coincide with the E and S cold fronts near the core. The overall temperature error along the "cold arm" is 10%. The higher southern temperatures near the CCD border have also correspondingly high errors (>~ 1 keV). There is a cold tail starting 2.3′ SW of the cluster's center that may extend S for more 5′(Tanaka et al. 2006).

In most simulations analyzed, AM06 found long lasting cold spiral arms coinciding with the cold fronts, sustained for many Gyrs after the sub-halo fly-by. In particular, their case for a DM perturber produces properties very similar to those observed in A496. In AM06 a pure dark matter halo with 1/5 of the mass of the main cluster flies-by with an impact parameter of 500 kpc and with closest approach at t∼1.37 Gyr (Fig 1d). Their simulations also seem to indicate the presence of milder cold fronts in the opposite side closer to cluster's core. These are clear predictions that are corroborated well in A496 and suggest strongly that a fly-by dark matter halo created the cold fronts in this cluster. A prediction of this model is the presence of massive DM halo in the outskirts of the cluster not accompanied by significant X-ray emitting gas. From the simulations, the position of that clump at epoch (t=2 Gyr, i.e., now) would be towards its apocenter at North, the same general direction of the main cold front (Dupke et al. 2006).

We acknowledge support from NASA through *Chandra* award number GO 4-5145X, NNG04GH85G and GO5-6139X and NASA grant NAG 5-3247.

Ascasibar, Y. & Markevitch, M., 2006, ApJ, in Press, astro-ph 0603246
Bialek, J. J., Evrard, A.E. & Mohr, J.J., 2002, ApJ, **578**, 9
Dupke, R. A. & White, R. E. III, 2003, ApJL, **583**, 13
Dupke, R.,, White, R. III, & Bregman J. 2006, ApJ, to be Submitted
Fabian, A. C., Sanders, J. S., Ettori, S., et al. 2001, MNRAS **321**, 33
Markevitch M. et al. 2000, ApJ, **541**, 542
Markevitch, M., Vikhlinin, A., & Mazzotta, P. 2001, ApJL, **562**, L153
Markevitch, M. et al. 2002, ApJL, 567, 27
Mazzotta, P., Markevitch, M., Vikhlinin, A., et al. 2001, ApJ, **555**, 205
Nagai, D. & Kravtsov, A. V., 2003, ApJ, **587**, 514
Tanaka, T., Kunieda, H., Hudaverdi, M., et al. 2006, PASJ, **58**, 703
Tittley, E. R., & Henriksen, M. 2003, ApJ, **563**, 673
Vikhlinin, A., Markevitch, M., & Murray, S. 2001, ApJ, **551**, 160

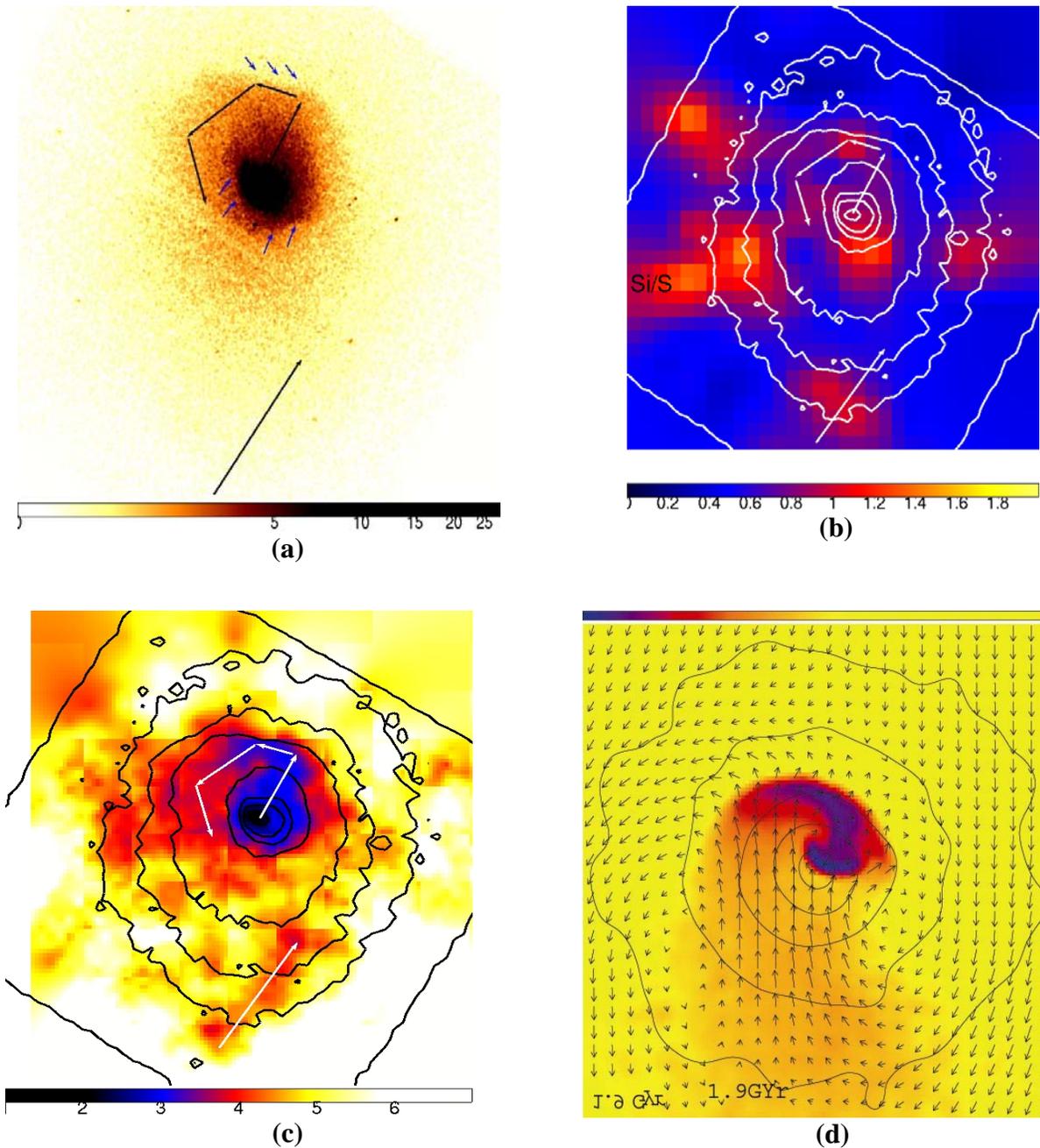

**Fig. 1.** **(a)** *Chandra* image of Abell 496. North is up. Blue arrows show the position of the northern (main), southern and eastern cold fronts. Black arrows near the center show the position of the low temperature and high SN II Fe mass fraction arm and the long arrow to the south shows the position of the cold "tail" in Figs a,b,c. **(b)** silicon to sulfur map. We also overlay the X-ray contours. The CCD border is also shown as the most external contour. Values shown for regions outside the CCD borders are an artifact of this type of code and should be neglected. The units are pixels and 1 pixel=0.5 arcsec. **(c)** Same as (b) but for gas temperatures **(d)** Zoom in of the core of simulated cluster 0.53 Gyr after the flyby of the dark matter halo, from AM06. Yellow is ~7–9 keV and blue 2 keV. DM density contours are overlaid and arrows indicate gas velocity (the longest corresponding to 500 km s−1). The size of the panel is 250 kpc, similar to the size of Acis-S3 CCD borders at the redshift of the cluster (~320 kpc). The figure has been flipped vertically to match the configuration of the cold front in A496. Figure best viewed in color.